\documentclass[a4paper,11pt]{JHEP}
\usepackage[T1]{fontenc}	
\usepackage{ae,aecompl,aeguill}		
\usepackage[ansinew]{inputenc}
\usepackage{amsmath}
\usepackage{amssymb}
\usepackage{slashed}
\usepackage{graphicx}
\usepackage{subfigure}

\newcommand{\be}{\begin{equation}}
\newcommand{\ee}{\end{equation}}
\newcommand{\beqa}{\begin{eqnarray}}
\newcommand{\eeqa}{\end{eqnarray}}

\newcommand\CLASS{{\tt CLASS}~}
\newcommand\CAMB{{\tt CAMB}~}
\newcommand\CMBFAST{{\tt CMBFAST}~}
\newcommand\CMBEASY{{\tt CMBEASY}~}
\newcommand\HALOFIT{{\tt HALOFIT}~}
\newcommand\CosmoMC{{\tt CosmoMC}~}
\newcommand\CosmoPMC{{\tt CosmoPMC}~}
\newcommand\MultiNest{{\tt MultiNest}~}

\preprint{CERN-PH-TH/2011-081, LAPTH-009/11} 

\title{The Cosmic Linear
  Anisotropy Solving System\\ (CLASS) I: Overview} \author{Julien Lesgourgues$^{a,b,c}$\vspace{.2cm}\\
  {$^a$}Institut de Th\'eorie des Ph\'enom\`enes Physiques,\\ \'Ecole
  Polytechnique F\'ed\'erale de Lausanne,\\ CH-1015, Lausanne,
  Switzerland.\vspace{.2cm}\\ {$^b$} CERN, Theory Division,\\ CH-1211
  Geneva 23, Switzerland.\vspace{.2cm}\\ {$^c$} LAPTh (CNRS -
  Universit\'e de Savoie), BP 110,\\ F-74941 Annecy-le-Vieux Cedex,
  France.}

\abstract{The Cosmic Linear Anisotropy Solving System ({\tt CLASS}) is a new
accurate Boltzmann code, designed to offer a more user-friendly and
flexible coding environment to cosmologists.
\CLASS is very structured, easy to modify,
and offers a rigorous way to control the accuracy of output quantities.
It is also incidentally a bit faster than other codes. In this
overview, we present the general principles of \CLASS and its basic
structure. We insist on the friendliness and flexibility aspects,
while accuracy, physical approximations and performances are discussed
in a series of companion papers.}

\begin{document} 

\section{Why a new code?}

There are several excellent Boltzmann codes publicly available on the
market, in particular \CMBFAST \cite{Seljak:1996is}, \CAMB
\cite{Lewis:1999bs} and \CMBEASY \cite{Doran:2003sy}. The second one is
still regularly maintained and upgraded, and offers lots of
functionality. What could we expect from a new Boltzmann code? Our
answer would be: user-friendliness, flexibility, accuracy and speed.

By friendliness, we mean: a code easy to understand, easy to compile
and to run on any platform, offering a convenient input interface, in
which input parameters could be passed in several ways.

Flexibility refers to the possibility to generalize the code to more
complicated cosmological scenarios, or to interface it with various
other programs (for parameter extraction, or for computing different
observables).

All accuracy parameters should be grouped in a unique place in order
to facilitate precision tuning. It should be possible to take the
decision to run with various well-identified and calibrated precision
levels, without needing to tune all accuracy parameters each time.

Since parameter extraction codes (e.g. \CosmoMC \cite{Lewis:2002ah},
\MultiNest \cite{Feroz:2008xx}, \CosmoPMC \cite{Kilbinger:2011bu})
typically require $10^4$ to $10^6$ calls to a Boltzmann code
(depending on the complexity of the model), any improvement in the
running time for a fixed precision setting would be appreciable.

Existing codes all contain several nice features, and brought decisive
progress with respect to their predecessors. \CMBFAST reduced
drastically the running time thanks to the line-of-sight integration
method, and included CMB lensing calculations for the first
time. \CAMB and \CMBFAST incorporate several refinements in terms of
structure of the code, lensing accuracy, approximation schemes, speed,
etc. Criticizing these codes would be totally out of place. But who
would claim that they are friendly and flexible? Anybody can download
them and learn how to run in one hour. But modifying these codes is
another story. Identifying the places where the modifications should
be performed is a very involved task. If the goal is to introduce
minimal modifications to the cosmological evolution, to interface the
code with another one, to compute another observable (e.g. lensing
spectra), a lot of work is already necessary. For implementing more
complicated changes (like e.g. adding dark matter species with a
non-trivial distribution or an interaction term), one should be -- or
become -- a dedicated expert.

Concerning accuracy, all current Boltzmann codes have been claimed to
be accurate at the 0.1\% level \cite{Seljak:2003th}. This is
marginally sufficient for a code supposed to analyze Planck data, for
which the error bars on temperature are really small for intermediate
angular scales (those with a very small cosmic variance and no
significant foreground contamination). With post-Planck CMB
experiments, cosmic shear surveys or 21cm data, even more accuracy
might be needed. But how can we be sure that existing Boltzmann code
are really accurate even at the 0.1\% level?  \CMBFAST and \CMBEASY
are not maintained anymore, their recombination algorithm is already
outdated. In ref. \cite{Hamann:2009yy}, these aspects were fixed for
the purpose of a \CMBFAST vs. \CAMB comparison, which revealed indeed
some very good agreement. But since the two codes have not been
developed in an independent way, their relative agreement does not
prove their absolute accuracy. Having an independent code would be the
only way to check this absolute accuracy, and to try to push the
precision further. \CAMB offers an option to tune some ``accuracy
boost parameters'', but the user does not know exactly to which
precision level a given setting corresponds. Without an independent
and up-to-date code, \CAMB's accuracy can only be calibrated ``with
respect to itself'', i.e. with respect to its
infinite-accuracy-parameter limit.

These issues of friendliness, flexibility and control of accuracy were
the main three motivations for developing a new code, the Comic Linear
Anisotropy Solving System (\CLASS)\footnote{available at {\tt
http://class-code.net}}. The fact that this code is also slightly
faster than its competitors is less important, but does not hurt.

\section{The \CLASS way}

\subsection{User friendliness}

\CLASS is written in C, a language familiar to most scientific
programmers, not very different from Fortran 90 but more
diffuse. Compiling \CLASS requires no specific version of the
compiler, no special package or library. In the {\tt Makefile}, the
compiling command is set by default to {\tt gcc -O4}, i.e. to the GNU
C compiler (installed on most computers) with optimization level 4. We
checked that the code is compatible with other compilers. After a {\tt
make class} command, the code should be ready to run.  On a multi-core
PC, one should try to run the code in parallel. This simply requires
the code to be compiled with an OpenMP option, e.g. {\tt gcc -O4
-fopenmp} (with the version 4.2 or higher of {\tt gcc}).  The OpenMP
flag usually has a different name for other compilers.

The code can be executed with a maximum of two input files, e.g.\\

\noindent 
{\tt ./class explanatory.ini chi2pl1.pre}\\

\noindent
The file with a {\tt .ini} extension is the cosmological parameter
input file, and the one with a {\tt .pre} extension is the precision
file. Both files are optional: all parameters are set to default
values corresponding to the ``most usual choices'', and are eventually
replaced by the parameters passed in the two input files. For
instance, if one is happy with default accuracy settings, it is enough
to run with {\tt ./class explanatory.ini}. Input files do not
necessarily contain a line for each parameter, since many of them can
be left to default value. The example file {\tt explanatory.ini} is
very long and somewhat indigestible, since it contains all possible
parameters, together with lengthy explanations. We recommend to keep
this file unchanged for reference, and to copy it in e.g. {\tt
test.ini}. In the latter file, the user can erase all sections in
which he/she is absolutely not interested (e.g., all the part on
isocurvature modes, or on tensors, or on non-cold species,
etc.). Another option is to create an input file from scratch, copying
just the relevant lines from {\tt explanatory.ini}.  For the simplest
applications, the user will just need a few lines for basic
cosmological parameters, one line for the {\tt output} entry (where
one can specifying which power spectra must be computed), and one line
for the {\tt root} entry (specifying the prefix of all output files).

The syntax of the input files is explained at the beginning of {\tt
explanatory.ini}.  Typically, lines in those files look like:

\vspace{0.5cm}

\noindent 
{\tt parameter1 = value1}\\
{\tt free comments}\\
{\tt parameter2 = value2 \# further comments}\\
{\tt \# commented\_parameter = commented\_value}\\

\noindent 
and parameters can be entered in arbitrary order. This is rather
intuitive. The user should just be careful not to put an ``{\tt =}''
sign not preceded by a ``{\tt \#}'' sign inside a comment: the code
would then think that one is trying to pass some unidentified input
parameter.

The syntax for the cosmological and precision parameters is the same.
It is clearer to split these parameters in the two files {\tt .ini}
and {\tt .pre}, but there is no strict rule about which parameter goes
into which file: in principle, precision parameters could be passed in
the {\tt .ini}, and vice-versa. The only important thing is not to pass
the same parameter twice: the code would then complain and not run.

The \CLASS input files are also user-friendly in the sense that many
different cosmological parameter bases can be used. This is made
possible by the fact that the code does not only read parameters, it
``interprets them'' with the level of logic which has been coded in
the {\tt input.c} module. For instance, the Hubble parameter, the photon
density, the baryon density and the ultra-relativistic neutrino density can be
entered as:

\vspace{0.5cm}

\noindent 
{\tt h = 0.7}\\
{\tt T\_cmb = 2.726 ~~~\# Kelvin units}\\
{\tt omega\_b = 0.02}\\
{\tt N\_eff = 3.04}\\

\noindent
(in arbitrary order), or as

\vspace{0.5cm}

\noindent 
{\tt H0 = 70}\\ {\tt omega\_g = 2.5e-5 ~~~\# g is the label for
photons}\\ {\tt Omega\_b = 0.04}\\ {\tt omega\_ur = 1.7e-5 ~~~\# ur is
the label for ultra-relativistic species}\\

\noindent
or any combination of the two. The code knows that for the photon
density, one should pass one (but not more than one) parameter out of
{\tt T\_cmb, omega\_g, Omega\_g} (where small omega's refer to
$\omega_i \equiv \Omega_i h^2$). It searches for one of these values,
and if needed, it converts it into one of the other two parameters,
using also other input parameters. For instance, {\tt omega\_g} will
be converted into {\tt Omega\_g} even if $h$ is written later in the
file than {\tt omega\_g}: the order makes no difference.  Lots of
alternatives have been defined. If the code finds that not enough
parameters have been passed for making consistent deductions, it will
complete the missing information with in-built default values. On the
contrary, if it finds that there is too much information and no unique
solution, it will complain and return an error. The code also writes a
{\tt root\_params.ini} file in output, in which the user can check
which non-default values have been used. Like in \CAMB, this file can
be used as an input file for another run.

In summary, the input syntax has been defined in such way that the
user does not need to think too much, and can pass his preferred set
of parameters in a nearly informal way.

Apart from easy compilation and easy input, user-friendliness is
achieved by sticking to well-known notations for all background and
perturbation quantities (in the synchronous and Newtonian gauge):
namely, those from Ma \& Bertschinger \cite{Ma:1995ey}. Most equations are
taken literally from this paper. The units are inverse mega-parsecs
for wavenumbers {\tt k} and for the Hubble rate, and mega-parsecs for
conformal time {\tt tau}.  For background densities $\rho_i$ and
pressure $p_i$, we use more unusual units. The goal is to be able to
write the Friedmann equation like
$$
{\tt H} = \left( \sum_{i=0}^{N} {\tt rho\_i} \right)^{1/2}~.
$$
Hence, everywhere in the code, the density {\tt rho\_i} stands for $[8
\pi G \rho_i / 3]$, in units of squared mega-parsecs (same for the
pressure {\tt p\_i}). The output CMB multipoles $C_l$'s are dimensionless
(unlike those of \CAMB which are in squared micro-Kelvins: hence, the
$C_l^{TT}$'s from \CLASS should be equal to those from \CAMB divided
by $[2.726 \times 10^6]^2$, if $T_{cmb}$ has been set to 2.726~K. 
Unless otherwise specified in the input file, the
units and the definition of output quantities are written in
comment lines at the top of each output file.

\subsection{Flexibility}

It is difficult to summarize in a few pages why we believe that \CLASS
is much easier to modify than other Boltzmann codes.  Broadly
speaking, we tried to achieve this goal by respecting three golden
rules:
\begin{itemize}
\item {\it no hard coding.} Any feature or equation which could be true in one cosmology
  and not in another one should not be written explicitly in the code,
  and should not be taken as granted in several other places.
  Discretization and integration steps should be defined automatically
  by the code, instead of being set to something which might be
  optimal for minimal models, and not sufficient for other
  ones. Physical relations which are likely to change in different
  cosmological scenarios should be localized in a small number of
  well-identified places in the code, that the user can access
  easily. Any step which depends of these relations should be
  performed automatically rather than in a model-dependent way. We
  provide several concrete examples below.
\item {\it clear structure.} The tasks of the code must be clearly
  separated: one module for the background evolution, another one for the
  thermodynamics evolution, another one for the perturbation evolution,
  etc. There should be no duplicate equations: a given physical
  assumption should be formulated in a single place.
\item {\it dynamical allocation of all indices.} On might be tempted to
  decide that in a given array, matrix or vector, a given quantity is
  associated with an explicit index value. However, when modifying the
  code, extra entries will be needed and will mess up the initial
  scheme; the user will need to study which index is associated to
  which quantity, and possibly make an error. All this can be avoided
  by using systematically a dynamical index allocation. This means
  that all indices remain under a symbolic form, and in each, run the
  code attributes automatically a value to each index. The user {\it
  never} needs to know this value.
\end{itemize}
As a result of these rules, the \CLASS files contain essentially no numbers, apart from:
\begin{itemize}
\item coefficient of physical equations, which can be transformed into
  input parameters if there is any reason that these coefficients
  could depend on the cosmology,
\item unit conversion factors and constants, defined in the {\tt
    include/*.h} files,
\item default values of input parameters, defined in the {\tt input} module.
\end{itemize}

\vspace{0.5cm}

Let us illustrate the flexibility of \CLASS through a few examples. 

\vspace{0.5cm}

{\bf Sampling steps.}  We need to integrate over time (or redshift) a
set of background, thermodynamical and perturbation equations. For
{\it background} equations, we use an adaptive Runge-Kutta integrator.
Such integrators need to be called several times over small time
steps.  The edge of these time steps define the discrete values of
time at which background quantities will be tabulated and stored, in
view of being interpolated in other modules. \CLASS infers the step
sizes automatically from the rate at which background equations
change. So, in a non-standard cosmological model where something
special would happen at a given time, the step size would
automatically decrease around that time. The same occurs with the {\it
  perturbation} equations when the user uses the same Runge-Kutta
integrator, which is only an option (the default integrator for
perturbations, called {\tt ndf15}, is described in
\cite{Blas:2011rf}). For {\it thermodynamical} quantities, the
equations are solved by the RECFAST module
\cite{Seager:1999bc,Scott:2009sz}, which we did not change much (apart
from small modifications leading to smoother results, as mentioned in
\cite{Lesgourgues:2011rg}). RECFAST chooses some step sizes for integrating
and sampling thermodynamical quantities during recombination. But at
low redshift, the evolution is modified according to some theoretical
ansatz for the enhanced free electron fraction $\Delta x_e(z)$ during reionization. This
function is chosen by the user. For whatever function $\Delta x_e(z)$, \CLASS
finds automatically the step sizes in redshift space which are
sufficient for capturing the evolution $x_e(z)$, given a dimensionless
tolerance parameter {\tt reionization\_sampling} (each step size is
given by $(d \ln x_e / dz)^{-1}$ times this parameter). A similar
logic is used by the code for defining the discrete times at which the
{\it source functions} $S(k, \tau)$ (needed to compute the observable
spectra) are sampled; or for defining the momentum bins in which the
{\it phase-space distribution} of each massive neutrinos and other non-cold
relics are calculated \cite{Lesgourgues:2011rh}.  As a result, the user is
free to change most physical equations (e.g. Friedmann and Einstein
equations, reionization function, phase-space distribution of
neutrinos) without ever needing to think about sampling issues. The
only step sizes which are not found automatically in \CLASS v1.0 are
those of wavenumbers $k$ (for source functions $S(k,\tau)$ and
transfer functions $\Delta_l(k)$) and of multipoles $l$ (for transfer
functions $\Delta_l(k)$ and harmonic spectra $C_l$'s), which are
either linear, logarithmic or a combination of the two. Those step
sizes are controlled by dimensionless parameters which can all be
tuned in the precision parameter file.

\vspace{0.5cm}

{\bf Reionization.} After computing the electron fraction $x_e(z)$
imposed by Helium/Hydrogen recombination using RECFAST, the code must
assume a given enhancement $\Delta x_e(z)$ of this function at low
redshift, accounting for reionization when the first stars form.  By
default, this function is often assumed to be a double step described
by two hyperbolic tangents, corresponding to Helium and Hydrogen
reionization. For each of these two steps, the user should pass input
parameters describing the mean redshift of the transition, its width,
its amplitude, and the redshift $z_{\rm max}$ above which we decide to
neglect reionization. In \CLASS, this function is not hard-coded, in
the sense that the user should just write inside the function {\tt
thermodynamics\_reionization\_function()} his favorite ansatz for
$\Delta x_e(z)$, depending on an arbitrary number of input
parameters. The code will do the rest automatically: finding
appropriate sampling steps, ensuring the continuity of the total
$x_e(z)$ around $z_{\rm max}$, finding the relation between the
reionization redshift and the optical depth (as in \CAMB, any of the two
can be passed in input, the other one will be inferred). Hence,
studying alternative reionization models is trivial with \CLASS.

\vspace{0.5cm}

{\bf Non-standard massive neutrinos and non-cold dark matter relics.}
Other publicly available Boltzmann codes include one or several
massive neutrino eigenstates, with one or several masses, but sharing
a unique phase space distribution $f(q)$ taken to be an exact
Fermi-Dirac. Changing this assumption (for studying non-standard
massive neutrinos with chemical potentials or non-thermal distortions,
light or heavy sterile neutrinos, warm dark matter candidates, etc.)
requires a large number of modifications to these codes, in which the
Fermi-Dirac distribution is hard-coded. The user interested in above
models should redefine $f(q)$ and its derivative $[d \ln f / d \ln q]$
in several places, eventually switch to a finer sampling of background
and perturbed quantities in $q$ space, redefine the relation between
mass and density, redefine some approximation schemes in the
relativistic/non-relativistic limits, etc. In \CLASS, for each
non-cold species $i$, the user should only modify in a {\it single
  place} the function $f_i(q)$. This function can depend on an
arbitrary number of parameters, which are easy to pass from the input
file. All the rest is done automatically: finding the sampling in
$q$-space using some dimensionless tolerance parameters; finding the
derivative $[d \ln f / d \ln q]$; finding the mass-to-density
relation; using approximation schemes, etc. The user is even free to
specify in the input file that for one or several of these species, the
distribution function should be read in a file rather than in a
function. This is useful for warm dark matter candidates, which
phase-space distribution function can be non-trivial and computable
only numerically. All these features are described in more detail in
\cite{Lesgourgues:2011rh}.

\vspace{0.5cm}

{\bf Adding new species in the code.}  \CLASS v1.0 includes the
following species with the following labels: photons ({\tt g},
compulsory); baryons ({\tt b}, compulsory); cold dark matter ({\tt
  cdm}, compulsory only in synchronous gauge); massless neutrinos and
other ultra-relativistic species ({\tt ur}, optional); arbitrary
number of massive neutrinos and other non-cold dark matter relics
({\tt ncdm}, optional); a non-perfect fluid with constant linear
equation of state and sound speed ({\tt fl}, optional); and a
cosmological constant ({\tt lambda}, optional). Including more species
is trivial thanks to the dynamical index allocation. The indices of
the above quantities are never hard-coded; for instance, in the vector
of all background quantities at a given time, the photon density reads
{\tt background[ba.index\_bg\_rho\_g]}, where ``{\tt index\_bg\_}''
means ``one index in the background vector'', ``{\tt rho\_g}'' means
``photon density'', and ``{\tt ba.}'' is necessary because this index
is defined inside the background structure, usually abbreviated as
{\tt ba} in the code. The user willing to code more species should
simply try to duplicate any reference to another species, and adapt
the names and the equations of motion. For instance, let us assume
that one wants to introduce another imperfect fluid with different
properties from {\tt fl}, and call it e.g. {\tt new}. The easiest
would be to search for all occurrences of the letters {\tt fl} in the
modules {\tt input, background, perturbations}, to duplicate them and
to replace {\tt fl} by {\tt new}. Then, the user can simply change the
equations of motion for {\tt new}, and define new input parameters
specifying the properties of the fluid in the same place where the
{\tt fl} parameters {\tt Omega\_fl}, {\tt w\_fl} and {\tt cs2\_fl} are
defined. By doing so, the user will be guided to define new variables
for the indices, e.g. {\tt index\_bg\_rho\_new} for the background
density, or {\tt index\_pt\_delta\_new} for the density
perturbations. All equations will be written in terms of these index
names, without needing to every worry about changes in the number of
equations, about explicit values of the indices, or about the order in
which quantities are arranged inside vectors and lists. The user will
also be guided to define a flag {\tt has\_new} which will be set to
``true'' inside the {\tt input module} only when the input value of
the {\tt new} density is non-zero. All references to the new fluid in
the code will be inside a condition ``{\tt if (has\_new == \_TRUE\_)
  \{...\}}''. This offers two advantages. First, when a species is not
present, the code ignores completely the lines referring to it. So, it does not slow down the
code to implement hundreds of new species and to run with only a few
of them being non-zero. Second, if the density of e.g. the {\tt new}
component is null, its index values are never assigned, but also never
needed by the code.

\vspace{0.5cm}

{\bf Adding new approximation schemes in the perturbation module.}
All Boltzmann codes use approximation schemes in order to reduce the
time spent in the perturbation integration. The most famous one is the
unavoidable tight-coupling approximation. In \CLASS, the notion of
approximation is formalized in such way that defining other
approximation schemes or skipping some of them is straightforward. Any
approximation scheme has a nickname: examples are the Tight-Coupling
Approximation {\tt tca}, the Ultra-relativistic Fluid Approximation
{\tt ufa}, the Radiation Streaming Approximation {\tt rsa} (all
described in \cite{Blas:2011rf}) or the Non-Cold Dark Matter Fluid
Approximation {\tt ncdmfa} (described in \cite{Lesgourgues:2011rh}).  Each of
these approximation are associated with ``status flags'' (e.g. {\tt
  tca\_on}, {\tt tca\_off}) and ``method flags'' (e.g.  {\tt
  tca\_class}, {\tt tca\_camb}, or just {\tt tca\_none} if this
approximation should never be used).  One of the ``method flags'' is
passed in input as a precision parameter, to state which equations
should be used when the approximation is turned on.  In the routine
{\tt subroutine\_approximations()}, one defines the condition under
which an approximation should be ``on'' or ``off''.  For instance, the
Ultra-relativistic Fluid Approximation should be ``on'' only if there
are non-relativistic species in the problem (this depends on the flag
{\tt has\_ur}), and if the ratio $k/aH$ exceeds a threshold value
defined as a precision parameters.  Before integrating perturbations
for a given wavenumber $k$, the code will automatically check how many
approximations need to be switched on or off for this wavenumber, and
will find each switching time $t_i$. The integration is then performed
inside each interval $[t_i, t_{i+1}]$ over which the approximation
scheme does not change. It means that at each switching time $t_i$,
the system of differential equations is redefined with a new size and
new initial values. With this logic, it is very simple to define new
approximations, or to modify the conditions under which an
approximation should be switched on or off, or finally to define a new
method corresponding to differens equations when the approximation is
switched on. Again, if one needs to code such a new approximation, the
easiest way is to search for all occurrence of one other nickname
(e.g. {\tt ufa}), to duplicate these occurrences, adapt the name, and
adapt the conditions and physical equations defining this
approximation.

\vspace{0.5cm}

{\bf Adding new source functions, new transfer functions, etc...}  In
the jargon of the code, a source function is just a function
$S(k,\tau)$ of wavenumber and time which can be inferred from
background, thermodynamical and perturbed quantities, and should be
stored during the execution of the {\tt perturbation} module
in order to be used by other modules. In fact, what remains in memory
after the integration of perturbed quantities is {\it only} a list of
tabulated source function. For CMB spectra, we need a temperature and
a polarization source function. For CMB lensing and cosmic shear, we
need another source, equal to the Newtonian gravitational potential:
$S(k,\tau)=\phi(k,\tau)$. For the matter power spectrum, we must store
one more source function equal to the matter density perturbation:
$S(k,\tau)=\delta_m(k,\tau)$. For outputting ``density transfer
functions'', i.e. the density of each species $i$ at time $\tau$
relative to the one at initial time, one should store several extra
source functions equal to $S(k,\tau)=\delta_i(k,\tau)/\delta_i(k,\tau_{ini})$
for each $i$. In the code, all source functions are defined with a
dynamical index and associated to a flag. For instance, if the
CMB spectra are needed (which is not always the case), the flag for
the temperature source {\tt has\_source\_t} will be set to true, and
any statement regarding this particular source will be inside a
condition ``{\tt if (has\_source\_t == \_TRUE\_) \{...\}}''.  Inside these
regions, the temperature source, which is one element in the vector of
all sources, will be referred as {\tt source[pt.index\_tp\_t]} where
{\tt index\_tp\_} means ``index for the type of source function'' and
{\tt t} means ``temperature''; this index is defined inside the
perturbation structure, usually called {\tt pt}. Once more, following
this logic, defining any other source function would be easy. The user
would need to track all occurrences of e.g. {\tt has\_source\_t}
or {\tt index\_tp\_t}, duplicate the corresponding lines, and replace
them with the name and definition of the new source functions.

\vspace{0.5cm}

{\bf Other examples.}  The very same logic based on flags and
dynamical indices is found everywhere in the code, in the
declaration of modes (scalars, tensors, etc.), of initial conditions
(adiabatic, different isocurvature types), of transfer functions
$\Delta_l(k)$, of observables (e.g. $C_l^{XX'}$ where $X$ stands for
temperature, polarization, lensing potential, etc.)... This means that
there are infinite ways to extend the code easily, using only the
search and the copy/paste commands, plus the many explanatory comments
written throughout the C files, and some basic theoretical knowledge
of cosmological perturbations.

\subsection{Control of accuracy}

The number of accuracy parameters in a Boltzmann code is surprisingly
large. In the \CAMB input file, only a few ``accuracy boost''
parameters and accuracy flags are visible; but they control many other
quantities. In \CLASS, we grouped all accuracy parameters within a
single structure, the ``precision structure'' declared in the {\tt
include/common.h} file.  As far as we remember, the user will not find
a single accuracy parameter declared and hard-coded locally in any
function or module. All these values are initialized in the {\tt
input} module and eventually overwritten by declarations in the input
file {\tt <...>.pre}. This is convenient, since the user willing to tune
the accuracy of \CLASS does not need to search for precision variables
dispersed throughout the code. The code is released together with a
few precision files which have been calibrated using a rigorous
process (see \cite{Lesgourgues:2011rg} for more details).

\subsection{Speed}

Speaking of the speed of the code only makes sense by referring to a
given precision level. The speed of \CLASS compared to that of \CAMB is
presented in \cite{Lesgourgues:2011rg} for a few accuracy settings and in the
minimal $\Lambda$CDM context. In this case, we found that \CLASS is
faster by a factor 2.5. This difference seems to result from the sum
of several improvements, rather than from a single trick. Progress
on the side of approximation schemes and integrators is described in
\cite{Blas:2011rf} for minimal $\Lambda$CDM, and in \cite{Lesgourgues:2011rh}
for models with massive neutrinos. Other improvements related to
numerical strategies and algorithms are impossible to summarize in a
paper, since the code was really written independently: inevitably,
the differences with respect to \CAMB are plethoric, and it is a
difficult task to identify which one are the most decisive for speeding
up the code.

The user should always try to compile with the most aggressive
optimization flags and with the OpenMP option, in order to be able to
run in parallel. We checked that if the number of OpenMP threads $N$
is not too large, the running time scales almost linearly with
$1/N$. If several runs are performed with the same precision
parameters, the code will find automatically that it can read
spherical Bessel functions in a file where they have been written
previously, in order to save a bit of time (a similar feature was
implemented in \CMBFAST). For usual precision settings, the computation
of Bessel functions is anyway very fast.  When running many occurrences
of \CLASS in a loop or in a parameter extraction code, one needs to
compute the Bessel function only once, and keep them in memory (this
is easy to achieve, as explained in the next section).

\section{Structure of the code}

\subsection{Directories}

After downloading \CLASS, one can see that files are split between the
following directories:
\begin{itemize} 
\item
{\tt source/} contains the C files for each \CLASS module, i.e. each
block containing some part of the physical equations and logic of
the Boltzmann code.
\item
{\tt tools/} contains purely numerical algorithms, applicable in any
context: integrators, simple manipulation of arrays (derivation,
integration, interpolation), Bessel function calculation, quadrature algorithms, parser, etc.
\item {\tt main/} contains the main module {\tt class.c} with the main
  routine {\tt class(...)}, to be used in interactive runs (but not
  necessarily when the code is interfaced with other ones).
\item
{\tt test/} contains alternative main routines which can be used to
run only some part of the code, to test its accuracy, to illustrate how
it can be interfaced with other codes, etc.
\item
{\tt include/} contains all the include files with a {\tt .h} suffix.
\item 
the root directory contains the {\tt Makefile} and some example of input files.
\end{itemize}

\subsection{Logic of each module}

All important modules are inside the {\tt source/} directory. The
structure of \CLASS is formalized in a rigorous way. It relies on 11
modules that we will call generically {\tt module\_i}, where $i$ is
supposed to be a label between 1 and 11. A module coincides with a C
file {\tt source/module\_i.c} and an include file {\tt
include/module\_i.h}. It is also associated with a structure that we
will call generically {\tt structure\_i}.  Finally, the module {\tt
module\_i.c} contains at least two functions {\tt
module\_i\_init.c(...)} and {\tt module\_i\_free.c(...)} which are the
essential ones.

The overall logic of \CLASS is summarized by the following two principles:
\begin{enumerate}
\item before the function {\tt module\_i\_init()} has been executed, {\tt
    structure\_i} contains some input parameters relevant for this
    module. These input parameter have been written inside the
    structure when the first module (the {\tt input.c} module) has
    been executed. For instance, in the case of the {\tt
    perturbation.c} module, the structure {\tt perturbation} contains
    information about which modes (scalar, tensors, ...) or which
    initial conditions (adiabatic, isocurvature...) are present.
\item after the module {\tt module\_i\_init()} has been executed, {\tt
    structure\_i} contains everything that other modules need to
    know. Intermediate quantities only computed internally in the
    module and not stored in {\tt structure\_i} are lost. For
    instance, after the execution of the {\tt perturbation.c} module,
    the structure {\tt perturbation} contains a table of source
    functions $S(k, \tau)$.
  \item when the various information stored in {\tt structure\_i} is
    not useful anymore, a call to {\tt module\_i\_free.c(...)} frees
    the memory dedicated to {\tt structure\_i}.
\end{enumerate}

The arguments of {\tt module\_i\_init()} should be: the precision
structure {\tt precision} which contains all accuracy parameters for
the whole code; the structures of the previous modules {\tt
structure\_1}, ..., {\tt structure\_(i-1)} which contain relevant
information; and the structure {\tt structure\_i}, which contains a
few input parameter before the execution of this function, and is
entirely filled afterward.  So, formally, executing the whole code
amounts in calling:

\vspace{0.5cm}

\noindent 
{\tt module\_1\_init(precision, structure\_1)}\\
{\tt module\_2\_init(precision, structure\_1,  structure\_2)}\\
...\\
{\tt module\_11\_init(precision, structure\_1,  structure\_2, ..., structure\_11)}\\
{\tt /* done, now free everything */ } \\
{\tt module\_11\_free(structure\_11)}\\
...\\
{\tt module\_1\_free(structure\_1)}\\

\noindent
The main routine in {\tt class.c} is therefore extremely compact (it
looks a bit longer than above only because of error management, as
explained below).

Each module contains more functions than just {\tt module\_i\_init()}
and {\tt module\_i\_free()}. These other functions are always named {\tt
  module\_i\_<...>()} (i.e., they always start with the name of the
module). They can be divided in two categories:
\begin{enumerate}
\item functions used only internally by each module. In the {\tt
    module\_i.c} file, these functions are always written after {\tt
    module\_i\_init()} and {\tt module\_i\_free()}.
\item functions which can be called by other modules. Typically, when
  a module $j$ needs quantities from an earlier module $i<j$, it can
  either read the information directly inside {\tt structure\_i}, or
  call a function inside {\tt module\_i.c} which knows how to read
  quantities in {\tt structure\_i} and returns a pointer to the needed
  information. For instance, after the {\tt background.c} module has
  been executed, the {\tt background} structure contains interpolation
  tables for all background quantities. If another module needs
  background quantities at a given conformal time $\tau$, instead of
  entering into the structure {\tt background}, this module can call
  the function {\tt background\_at\_tau(tau, ...)}, which will return
  the desired quantities. Functions called by other modules are always
  written before {\tt module\_i\_init()} in the {\tt module\_i.c}
  file, since they are those deserving more visibility.
\end{enumerate}

\subsection{The \CLASS backbone}

Executing \CLASS amounts to execute eleven functions of the type {\tt
module\_i\_init()}, for the following eleven modules:

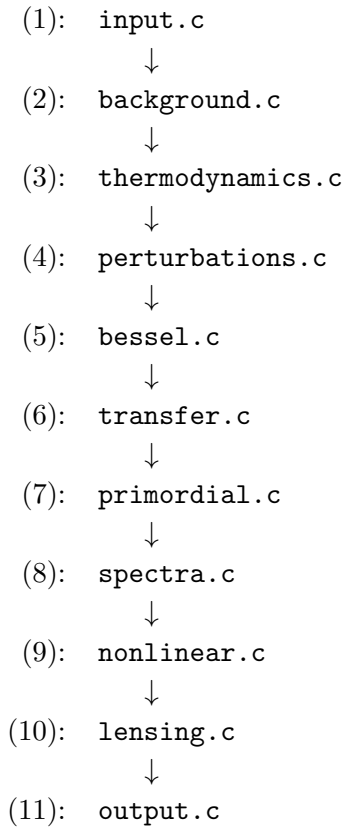
\vspace{0.5cm}

\begin{tabular}{rl}
(1): & {\tt input.c} \\
&$~~~~\downarrow$\\
(2): & {\tt background.c} \\
&$~~~~\downarrow$\\
(3): & {\tt thermodynamics.c} \\
&$~~~~\downarrow$\\
(4): & {\tt perturbations.c} \\
&$~~~~\downarrow$\\
(5): & {\tt bessel.c} \\
&$~~~~\downarrow$\\
(6): & {\tt transfer.c} \\
&$~~~~\downarrow$\\
(7): & {\tt primordial.c} \\
&$~~~~\downarrow$\\
(8): & {\tt spectra.c} \\
&$~~~~\downarrow$\\
(9): & {\tt nonlinear.c} \\
&$~~~~\downarrow$\\
(10): & {\tt lensing.c} \\
&$~~~~\downarrow$\\
(11): & {\tt output.c}\\
\end{tabular}

\vspace{0.2cm}

\noindent 
So, the {\tt main()} function is identical to the scheme described in the
previous section, with the above module names, called in the above
order.  This order is defined by the fact that a function {\tt
  module\_i\_init()} needs the structures {\tt structure\_j} for $j<i$ to
be already filled. There is actually a little bit of freedom.  For
instance, the only argument of the {\tt bessel\_init()} function is the
{\tt precision} structure, initialized in the {\tt input.c} module and
passed to all {\tt module\_i\_init()} functions; as well as the {\tt
  bessel} structure, for which input parameter are initialize as usual
in the {\tt input.c} module, and the rest is filled within the {\tt
  bessel\_init()} function. The {\tt bessel} structure is used only by
the {\tt transfer.c} module. This means that {\tt bessel\_init()} could
actually be called at any time between {\tt input\_init()} and {\tt
  transfer\_init()}. But there is not as much flexibility for other
modules, since most of them really use the structure of the previous module
$(i-1)$, while their own structure is requested by the next module $(i+1)$.

This structure implements a very clear separation of the physical
tasks.  The user can easily know where is the region that he
eventually needs to modify.  Moreover, somebody can only be interested
in computing the background evolution, or the thermodynamical
evolution, or the transfer functions. In this case, instead of using
the full sequence of modules, it is possible to execute only the first
few modules. The alternative {\tt main()} functions contained in the
{\tt test/} directory offer such examples.

\subsection{Quick glance at each module}

\begin{itemize}
\item
{\tt input.c}, the first module, is a bit different from the other ones. It
is the only module not associated with a structure (an {\tt input} structure
would have been useless), and it is also the only module which has two {\tt
\_ini()} functions. The first one, {\tt
input\_init\_from\_arguments()}, is the first function called by \CLASS
when the code is run interactively using the main routine in {\tt
main/class.c}. It reads a maximum of two arguments, the {\tt <...>.ini}
input file and the {\tt <...>.pre} precision file. It sets all input and
precision parameters to default values, and then eventually replace
some of them with the values indicated in the file(s) (but the same
parameter cannot be reset twice in the files, otherwise the code will
complain). As we said earlier, the fact that input parameters go into
{\tt <...>.ini} and precision parameters into {\tt <...>.pre} is not a
strict rule: in practise the code just considers the sum of the two
files, and everything could be passed in a single file. The second
function, {\tt input\_init()}, is the first function to be called when the
code is interfaced with another one. In this case, input parameters
are not read from the arguments of the function, but from a structure
called {\tt file\_content} which contains the same information as the
files, in roughly the same format. So, if \CLASS is to be run
e.g. within a parameter extraction code, this code should write
input/precision parameter inside a {\tt file\_content} structure, and
then call {\tt input\_init()}. Actually, the role of the function {\tt
input\_init\_from\_arguments()} is to convert the content of the input
files into a structure {\tt file\_content}, and then to call {\tt
input\_init()}. The list of arguments for these two functions is long:
indeed, they need a pointer towards each of the structures of all
other modules, in order to initialize them. Since {\tt input.c} is the
only module not having its own structure, it is also the only one
without a {\tt \_free()} function.

\item
{\tt background.c} simply solves the background equations (in
particular, the Friedmann equation), and stores into its structure {\tt
background} an interpolation table for all background quantities as a
function of time. Other modules often call the function {\tt
background\_at\_tau()} which returns these quantities interpolated at
some conformal time $\tau$, and the function {\tt
background\_tau\_of\_z()} which converts a value of redshift into a
value of conformal time. This module also stores in the {\tt
background} structure useful background-related quantities like the
age of the universe.

\item {\tt thermodynamics.c} solves for the thermodynamical evolution
  with RECFAST, corrects it for reionization, and stores into its
  structure {\tt thermo} an interpolation table for all thermodynamical
  quantities as a function of redshift. Other modules often call the
  function {\tt thermodynamics\_at\_z()} which returns these
  quantities interpolated at some redshift $z$.  This modules also stores
  in the {\tt thermo} structure useful quantities like the
  recombination time, reionization time, reionization optical depth if
  it was not passed in input, reionization redshift if it was not
  passed in input, sound horizon at recombination, etc.

\item {\tt perturbation.c} solves the evolution of all perturbations,
and stores the source functions $S(k,\tau)$ in a table inside its
structure {\tt perturbs}. Note that when perturbations are integrated
for a given wavenumber, background and thermodynamical quantities are
not evolved another time: they are interpolated using the functions
{\tt background\_at\_tau()} and {\tt thermodynamics\_at\_z()}, in
order to enforce a clear separation of the tasks. As a result, the
Friedmann equation only appears in a single line in the {\tt
background.c} module.

\item {\tt bessel.c} computes spherical Bessel functions and stores them 
in its structure {\tt bessels}. We already mentioned that this module
could also be called earlier, just after the input module. Actually,
when \CLASS is to be run several time within e.g. a parameter
extraction code, always with the same accuracy parameters, one should
keep the bessel functions in memory and not call this module each time. In
practise, this can be done in the following way: call once {\tt
input\_init()} and {\tt bessel\_init()}; then, for each new model,
call again {\tt input\_init()} and all other {\tt
\_init()} functions in each module, but not {\tt bessel\_init()}
again. Free all structures by calling in reverse order all {\tt
\_free()} functions, but not {\tt bessel\_free()}. Finally, when
everything is finished, do a final call of {\tt bessel\_free()}.

\item {\tt transfer.c} computes transfer functions $\Delta_l(k)$ (by convolving source functions and Bessel functions), and stores them in its structure {\tt transfers}.

\item {\tt primordial.c} computes the primordial power spectra (for
each mode and initial condition) and stores them in its structure {\tt
primordial}. In \CLASS v1.0, only simple analytical formulas are
implemented in this module, but in the future, it will be able to call
an inflation simulation module if requested.

\item {\tt spectra.c} computes observable power spectra out of source
  functions, transfer functions and primordial spectra, and stores
  them in its structure {\tt spectra}.

\item {\tt nonlinear.c} gives an estimate of the non-linear version of
  the previous spectra, according to some scheme chosen by the user,
  and stores them in the structure {\tt nonlinear}. Very soon, a
  renormalization scheme will be available, as well as other methods.

\item {\tt lensing.c} computes lensed temperature and polarization CMB
  spectra, using the unlensed spectra and the CMB lensing potential
  spectrum, and stores them in the structure {\tt lensing}. \CLASS
  computes the lensed CMB spectra from all-sky correlation
  functions~\cite{Challinor:2005jy}, i.e with the same method as
  \CAMB, but with a different numerical implementation written by
  S. Prunet, based on quadrature weigths.

\item {\tt output.c} just writes the output in some files. This module
  does not need to be called when \CLASS is used inside another code,
  e.g. a parameter extraction code. Otherwise, the function {\tt
    output\_ini()} is the last one called in the main routine, before
  freeing structures. Actually, all fields in the {\tt output}
  structure can be freed at the end of {\tt output\_init()}, so there
  is no need to call {\tt output\_free()}, which is not even defined.
\end{itemize}

\subsection{Error management}

Error management is based on the fact that all functions are defined
as integers returning either {\tt \_SUCCESS\_} or {\tt
  \_FAILURE\_}. Before returning {\tt \_FAILURE\_}, they write an
error message in the structure of the module to which they belong. The
calling function will read this message, append it to its own error
message, and return a {\tt \_FAILURE\_}; and so on and so forth, until
the main routine is reached. This error management allows the user to
see the whole nested structure of error messages when an error has
been found. The structure associated to each module contains a field
for writing error messages, called {\tt
  structure\_i.error\_message}. So, when a function from a module $i$
is called within module $j$ and returns an error, the goal is to write
in {\tt structure\_j.error\_message} a local error message, and to
append to it the error message in {\tt
  structure\_i.error\_message}. These steps are implemented in a macro
{\tt class\_call()}, used for calling whatever function:

\vspace{0.5cm}

\noindent
{\tt class\_call(module\_i\_function(...,structure\_i),}\\
\mbox{ }~~~~~~~~~~~~~~~~{\tt structure\_i.error\_message,} \\ 
\mbox{ }~~~~~~~~~~~~~~~~{\tt structure\_j.error\_message)}\\
    
\noindent
So, the first argument of {\tt call\_call()} is the function we want
to call; the second argument is the location of the error message
returned by this function; and the third one is the location of the
error message which should be returned to the higher level. The user
will find in {\tt include/common.h} a list of additional macros, all
starting by {\tt class\_...()}, which are all based on this logic (for
instance, the macro {\tt class\_test()} offers a generic way to return
an error in a standard format if a condition is not fulfilled). A
typical error message from \CLASS looks like:

\vspace{0.5cm}

\noindent
{\tt Error in module\_j\_function1\\
=> module\_j\_function1 (L:340) : error in module\_i\_function2(...)\\
=> module\_i\_function2 (L:275) : error in module\_k\_function3(...)\\
...\\
=> module\_x\_functionN (L:735) : your choice of input parameter blabla=30
is not consistent with the constraint blabla<1\\}

\noindent
where the {\tt L}'s refer to line numbers in each file. These error
messages are very informative, and are built almost entirely
automatically by the macros. For instance, in the above example, it
was only necessary to write inside a {\tt class\_test()} macro the
words {\tt `your choice of input parameter blabla = \%g is not
consistent with the constraint blabla < \%g',blabla, blabla\_max}. All
the rest was added step by step by the various {\tt
class\_call()} macros.

\section{What is already there, and what is next?}

In terms of cosmological species, as already mentioned previously,
\CLASS v1.0 includes photons ({\tt g}, compulsory); baryons ({\tt b},
compulsory); cold dark matter ({\tt cdm}, compulsory only in
synchronous gauge); massless neutrinos and other ultra-relativistic
species ({\tt ur}, optional); arbitrary number of massive neutrinos
and other non-cold dark matter relics ({\tt ncdm}, optional); a
non-perfect fluid with constant linear equation of state and sound
speed ({\tt fl}, optional); and a cosmological constant ({\tt lambda},
optional). Initial conditions can be an arbitrary mixture of an
arbitrary number of arbitrarily correlated adiabatic and isocurvature
modes. For each auto-correlation and cross-correlation spectra, the
user can enter an amplitude, a tilt and a running. Non-cold dark
matter relics can be tuned with a variety of options described in the
{\tt explanatory.ini} input file and illustrated in \cite{Lesgourgues:2011rh}:
free masses, temperatures, chemical potentials for flavor species,
mixing angles (used to go from the flavor basis to the mass basis),
degeneracy parameters, and as mentioned above, possibility to change
the phase-space distribution, using an analytical expression or
tabulated values in a file.

The code includes tensor modes, which have not been tested thoroughly
in v1.0 like the scalar ones (progress is still expected in this
direction). It computes the CMB anisotropy spectra for the products
$TT$, $TE$, $EE$, $BB$, as well as the auto-correlation and
cross-correlation spectra involving the CMB lensing potential
$\phi$. It computes also the lensed anisotropy spectra. It can output
the matter power spectrum, today or for an arbitrary list of redshifts
passed by the user; same for the density transfer functions of each
species.

Coding the Newtonian gauge is almost finished, but only
the synchronous gauge is operational in version 1.0.

The next steps will consist in improving the way to deal with
tensors, in making the Newtonian gauge operational, and in coding
open/closed models. We wish to release interfaces between \CLASS and
the parameter extraction codes \CosmoMC, \CosmoPMC and \MultiNest. It
would be easy to add more general parametrizations for the
reionization history, and to adapt to \CLASS our \CAMB inflationary
module \cite{Lesgourgues:2007gp,Lesgourgues:2007aa}. We will release
very soon a renormalization algorithm embedded in the {\tt
nonlinear.c} module, in which we will also implement non-linear
approximations like \HALOFIT \cite{Smith:2002dz}. It will finally be
easy to add more observables, like e.g. correlation functions and
power spectra for cosmic shear survey.

\section*{Acknowledgments}

We wish to thank Fran\c{c}ois Bouchet for inspiring this project, as
well as Martin Bucher, Damien Girard, Jan Hamann and Alain Riazuelo
for useful input in its earliest stages. Throughout the realization of
\CLASS, Karim Benabed and Simon Prunet provided essential help and
support. Benjamin Audren and Simon Prunet coded respectively the
renormalization and lensing algorithms. In the past ten months, Thomas
Tram boosted the last coding steps without sparing time and
efforts. Enthusiastic moral support from several friends and
colleagues was very much appreciated.



\begin{thebibliography}{99}

\bibitem{Seljak:1996is}
  U.~Seljak, M.~Zaldarriaga,
  Astrophys.\ J.\  {\bf 469 } (1996)  437-444.
  [astro-ph/9603033].

\bibitem{Lewis:1999bs}
  A.~Lewis, A.~Challinor, A.~Lasenby,
  Astrophys.\ J.\  {\bf 538 } (2000)  473-476.
  [astro-ph/9911177].

\bibitem{Doran:2003sy}
  M.~Doran,
  JCAP {\bf 0510 } (2005)  011.
  [astro-ph/0302138].

\bibitem{Lewis:2002ah}
  A.~Lewis, S.~Bridle,
  Phys.\ Rev.\  {\bf D66 } (2002)  103511.
  [astro-ph/0205436].

\bibitem{Feroz:2008xx}
  F.~Feroz, M.~P.~Hobson, M.~Bridges,
  [arXiv:0809.3437 [astro-ph]].

\bibitem{Kilbinger:2011bu}
  M.~Kilbinger, K.~Benabed, O.~Cappe, J.~-F.~Cardoso, G.~Fort, S.~Prunet, C.~P.~Robert, D.~Wraith,
  [arXiv:1101.0950 [astro-ph.CO]].

\bibitem{Seljak:2003th}
  U.~Seljak, N.~Sugiyama, M.~J.~White, 1, M.~Zaldarriaga,
  Phys.\ Rev.\  {\bf D68 } (2003)  083507.
  [astro-ph/0306052].

\bibitem{Hamann:2009yy}
  J.~Hamann, A.~Balbi, J.~Lesgourgues, C.~Quercellini,
  JCAP {\bf 0904 } (2009)  011.
  [arXiv:0903.0382 [astro-ph.CO]].

\bibitem{Ma:1995ey}
  C.~P.~Ma and E.~Bertschinger,
  Astrophys.\ J.\  {\bf 455} (1995) 7
  [arXiv:astro-ph/9506072].

\bibitem{Blas:2011rf}
  D.~Blas, J.~Lesgourgues, T.~Tram,
  ``The Cosmic Linear Anisotropy Solving System (CLASS) II: Approximation schemes,''
  [arXiv:1104.2933 [astro-ph.CO]].

\bibitem{Lesgourgues:2011rg}
  J.~Lesgourgues,
  ``The Cosmic Linear Anisotropy Solving System (CLASS) III: Comparision with CAMB for LambdaCDM,''
  [arXiv:1104.2934 [astro-ph.CO]].

\bibitem{Seager:1999bc}
  S.~Seager, D.~D.~Sasselov, D.~Scott,
  Astrophys.\ J.\  {\bf 523 } (1999)  L1-L5.
  [astro-ph/9909275].

\bibitem{Scott:2009sz}
  D.~Scott and A.~Moss,
  arXiv:0902.3438 [astro-ph.CO].

\bibitem{Lesgourgues:2011rh}
  J.~Lesgourgues, T.~Tram,
  ``The Cosmic Linear Anisotropy Solving System (CLASS) IV: Efficient implementation of non-cold relics,''
  [arXiv:1104.2935 [astro-ph.CO]].

\bibitem{Challinor:2005jy}
  A.~Challinor, A.~Lewis,
  Phys.\ Rev.\  {\bf D71 } (2005)  103010.
  [astro-ph/0502425].

\bibitem{Lesgourgues:2007gp}
  J.~Lesgourgues, W.~Valkenburg,
  Phys.\ Rev.\  {\bf D75 } (2007)  123519.
  [astro-ph/0703625 [ASTRO-PH]]

\bibitem{Lesgourgues:2007aa}
  J.~Lesgourgues, A.~A.~Starobinsky, W.~Valkenburg,
  JCAP {\bf 0801 } (2008)  010.
  [arXiv:0710.1630 [astro-ph]].

\bibitem{Smith:2002dz}
  R.~E.~Smith {\it et al.} [ The Virgo Consortium Collaboration ],
  Mon.\ Not.\ Roy.\ Astron.\ Soc.\  {\bf 341 } (2003)  1311.
  [astro-ph/0207664].

\end{thebibliography}
\end{document}